\documentclass[12pt]{iopart}
\usepackage{iopams}
\usepackage{bm}
\usepackage{graphicx,subfigure}
\usepackage{hyperref}
\usepackage{color}
\usepackage{graphics}
\usepackage{epsfig}

\hypersetup{
pdfauthor = {Daniel, Mucio, Fernanda, Christopher},
pdftitle = {Pressure induced BEC-BCS crossover in multi-band superconductors, \today.},
}
\begin{document}
\title[Pressure induced BEC-BCS crossover in multi-band superconductors]{Pressure induced BEC-BCS crossover in multi-band superconductors}
%

\author{D Reyes$^1$, M A Continentino$^2$, F Deus$^3$, C Thomas$^4$}

\address{$^1$ Instituto Militar de Engenharia - Pra\c{c}a General Tib\'{u}rcio, 80, 22290-270, Praia Vermelha, Rio de Janeiro, Brazil}
\ead{daniel@cbpf.br}
\address{$^2$ Centro Brasileiro de Pesquisas F\'{\i}sicas - Rua Dr. Xavier Sigaud, 150-Urca, 22290-180, RJ-Brazil}
\address{$^3$ Universidade do Estado do Rio de Janeiro, Faculdade de Tecnologia, Departamento de Matem\'{a}tica, F\'{\i}sica e Computa\c{c}\~{a}o,
Rodovia Presidente Dutra km 298, 27537-000, Resende, RJ, Brazil}
\address{$^4$ Departamento de F\'isica, Universidade Federal Rural do Rio de Janeiro, 23897-000, Serop\'edica, Rio de Janeiro, Brazil}
\date{\today}
\begin{abstract}
Superconductivity in strongly correlated systems is a remarkable phenomenon that attracts a huge interest.  The study of this problem is relevant for materials as the  high $T_c$ oxides,  pnictides and heavy fermions. These systems also have in common the existence of electrons of several orbitals that coexist at a common Fermi-surface. In this paper we study the effect of  pressure, chemical or applied on multi-band superconductivity. Pressure varies the atomic distances and consequently the overlap of the wave-functions in the crystal. This rearranges the electronic structure that we model including a pressure dependent hybridization between the bands. We consider the case of two-dimensional systems in a square lattice with inverted bands. We study the conditions for obtaining a pressure induced superconductor quantum critical point and show that hybridization, i.e., pressure can induce a BCS-BEC crossover in multi-band systems even for moderate interactions. We briefly discuss the influence of the symmetry of the order parameter in the results.
\end{abstract}

\maketitle
\ioptwocol

\section{Introduction}\label{secint}

The study of superconductivity in strongly correlated electron systems (SCES) is one of the most exciting areas in condensed matter physics~\cite{fese}. It encompasses many interesting materials including those that are most promising for practical applications, as high $T_c$ materials like  the cuprates and the family of iron pnictides~\cite{multi,3high,rafa1}.  Also  heavy fermion materials that in spite of their small critical temperature arise an enormous interest for presenting clear and unambiguous evidence  of superconductivity associated to a magnetic quantum critical point~\cite{gil}. In all these superconductors,  electronic correlations play an important role and  their relevance for the superconducting properties is fully recognized.  These systems however have in common another important characteristic. They are multi-band superconductors with  electrons from different atomic orbitals coexisting at the Fermi surface. These orbitals overlap in real space and are mixed by the crystalline potential giving rise to bands with hybrid character. 
Hybridization  depends  on the atomic distances and can vary as a function of external pressure or doping. This sensitivity of the mixing of atomic orbitals and consequently of the hybridized bands  to pressure or doping provides a useful control parameter that can be tuned and used to explore the phase diagram of multi-band superconductors~\cite{padilha, ramos, fang, continentino, li, heron}. It is this aspect of hybridization that we explore in this paper. We consider a two-band system each formed by electrons in the same orbitals  possibly hybridized with other states~\cite{raghu,flores}.  We then add to this two-band problem a new hybridization term that we will vary with the intention of simulating the effects of external and chemical pressure in the electronic structure. Furthermore we consider an inter-band interaction, of the Josephson type that transfers pairs of electrons  from one hybrid band to another~\cite{suhl}. This type of interaction, that will be responsible for superconductivity is  generally considered the most relevant in  multi-band superconductors~\cite{rafa1, suhl}. 

Another important feature of the systems we are considering is the presence of strong interactions responsible for the formation of Cooper pairs. In high-$T_c$ materials the strong coupling nature of the superconductivity is rather evident. However, this is  also the case for heavy fermions, as becomes clear when comparing the   $T_c$ of these systems with their very low effective Fermi temperatures~\cite{gil,spalek}. For this case of strong interactions it will be necessary to extend the usual BCS approach to allow for variation of the chemical potential as is usually done in the study of the BCS-BEC crossover~\cite{nozieres, samelos, carter, jre, babaev, adachi, nikolic}.

In this paper we present a detailed study of superconductivity in multi-band systems as we vary the hybridization and the electronic occupation of the bands for arbitrary strengths of the  interactions. We consider in detail the case of attractive interactions. Since we determines both the order parameter and chemical potential self-consistently we show that increasing hybridization promotes a BCS-BEC crossover  similar to increasing the strength of the interactions.
The results we obtain are of general interest. Our goal is not to provide a detailed study of a given material, but to investigate general features of the phase diagram of multi-band superconductors as they are subjected to applied pressure and the  occupation of the electronic bands is changed. We also discuss the possibility of superconducting quantum critical points (SQCP) and of a BCS-BEC crossover appearing in these systems as a function of external pressure.

\section{The model}

We consider a two-dimensional two-band lattice model with the Hamiltonian  given by,
\begin{eqnarray}\label{hamilt}
  \mathcal{H}&=&\sum_{\bf{k} \sigma}\! \left( \epsilon_{0\bf{k}}^{a} a_{\bf{k} \sigma}^{\dag}  a_{\bf{k} \sigma}\!+\!\epsilon_{0\bf{k}}^{b} b_{\bf{k} \sigma}^{\dag} b_{\bf{k} \sigma}\!\right) \nonumber \\
&+&
 V \sum_{\bf{k} \sigma}\! \left( \!a_{\bf{k} \sigma}^{\dag} b_{\bf{k} \sigma} \!+\! b_{\bf{k} \sigma}^{\dag} a_{\bf{k} \sigma}\!\right) \! -\! \mu \sum_{\bf{k}, \sigma} \left( n_{\bf{k} \sigma}^a + n_{\bf{k} \sigma}^b \right) \nonumber \\
&-&J \sum_{\bf{k} \bf{k}^{\prime}} \left( a_{\bf{k} \uparrow}^{\dag} a_{- \bf{k} \downarrow}^{\dag} b_{-\bf{k}^{\prime} \downarrow} b_{\bf{k}^{\prime} \uparrow} +b_{\bf{k} \uparrow}^{\dag} b_{- \bf{k} \downarrow}^{\dag} a_{-\bf{k}^{\prime} \downarrow} a_{\bf{k}^{\prime} \uparrow} \right), \nonumber \\
\end{eqnarray}
where $ a_{\bf{k} \sigma}^{\dagger}$ and $ b_{\bf{k} \sigma}^{\dagger}$ are creation operators for fermions in the $a$ and $b$ bands, respectively. These bands are described by the dispersion relations, $\epsilon_{\bf{k}}^{a}$ and $\epsilon_{\bf{k}}^{b}$, for $a$ and $b$-quasi-particles in an obvious notation. In the spirit of our discussion in the Introduction these are generic bands that take into account the atomic orbitals and the mixing between them. They can also represent bands with electron and hole character, as in the case of the $Fe$ superconductors~\cite{fese,andrey}. Besides, they may be distinguished by their effective masses; one band has lighter quasi-particles, the other contains more localized, heavier fermionic excitations. The $\langle i,j \rangle$ refer to lattice sites and $\sigma$ denotes the spin of the quasi-particles. The two  electronic bands can further hybridize through  a $k$-independent matrix element $V$~\cite{D16} that can be tuned by external parameters such as pressure permitting the exploration of the phase diagram and quantum phase transitions of the model. The quantity $J$ represents an inter-band interaction between the quasi-particles in the two-bands. This type of coupling is generally referred as of the {\it Josephson type} since it transfers pairs of quasi-particles between the bands~\cite{suhl}. Finally $\mu$ is the chemical potential that fixes the total number of electrons $n_{\mathrm{tot}}$ in the system.
Neglecting any form of magnetic order, the model given by (\ref{hamilt}) can be decoupled in the Cooper channel, at the mean-field level to yield,
\begin{eqnarray}\label{hamiltbcs}
  \mathcal{H} &=& \sum_{\bf{k} \sigma}\! \left( \epsilon_{0\bf{k}}^{a} a_{\bf{k} \sigma}^{\dag}  a_{\bf{k} \sigma}+\epsilon_{0\bf{k}}^{b} b_{\bf{k} \sigma}^{\dag} b_{\bf{k} \sigma}\!\right) \nonumber \\
  &+& V \sum_{\bf{k} \sigma}\! \left( \!a_{\bf{k} \sigma}^{\dag} b_{\bf{k} \sigma} \!+\! b_{\bf{k} \sigma}^{\dag} a_{\bf{k} \sigma}\!\right) \! -\! \mu \sum_{\bf{k}, \sigma} \left( n_{\bf{k} \sigma}^a+n_{\bf{k} \sigma}^b \right) \nonumber \\
&-& \sum_{\bf{k}}  \left(\Delta_{b} a_{- \bf{k} \downarrow} a_{\bf{k} \uparrow} + \Delta_{b}^{*} a_{\bf{k} \uparrow}^{\dag} a_{- \bf{k} \downarrow}^{\dag}\right) \nonumber \\  &-&  \sum_{\bf{k}} \left(\Delta_{a} b_{- \bf{k} \downarrow} b_{\bf{k} \uparrow} + \Delta_{a}^{*} b_{\bf{k} \uparrow}^{\dag} b_{- \bf{k} \downarrow}^{\dag}\right),
\end{eqnarray}
with
\begin{eqnarray}
\label{deltas}
&& |\Delta_{a}| = J \sum_{\bf{k}}  \langle a_{\bf{k} \uparrow}^{\dag} a_{-\bf{k} \downarrow}^{\dag} \rangle, \\
&& |\Delta_{b}| = J \sum_{\bf{k}}  \langle b_{\bf{k} \uparrow}^{\dag} b_{- \bf{k} \downarrow}^{\dag} \rangle,
\end{eqnarray}
where $\Delta_{a,(b)}=|\Delta_{a,(b)}|e^{\imath \phi_{a,(b)}}$ are  complexes  superconductor order parameters with amplitude $|\Delta_{a,(b)}|$ and phase $\phi_{a,(b)}$. The  dispersion of the  $a$ and $b$ bands are taken in a 2D square lattice with nearest neighbor hopping and given by, $\epsilon_{0\bf{k}}^{a}=-2t(\cos (k_x \mathtt{a})+\cos (k_y \mathtt{a}))$  and $\epsilon_{0 \bf{k}}^{b}=\epsilon_{0} + \alpha\epsilon_{0 \bf{k}}^a$, respectively. The quantity $t$ is an effective hopping that includes the transfer of electrons within the same orbital and the mixing between different orbitals.  The quantity $\epsilon_{0}$  is the relative shift between the two bands and $\alpha<1$ takes into account the ratio of the effective masses of the quasi-particles in different bands.

\subsection{Spectrum of Excitations}
\label{spectrum}
The energy of the quasi-particle excitations in the superconducting phase of the model described by (\ref{hamiltbcs}) can be obtained exactly. For this purpose, we use the equations of motion for the normal and anomalous Green's function~\cite{D16,Daniel2007}. The poles of the Green's functions yield the spectrum of excitations in the superconducting phase. The energies of these modes are given by,
\begin{eqnarray}
\label{bogoliubons}
&& \omega_{1} = \sqrt{A_{\bf{k}} + \sqrt{A_{\bf{k}}^{2} - B_{\bf{k}}}} = - \omega_{3}, \\
&& \omega_{2} = \sqrt{A_{\bf{k}} - \sqrt{A_{\bf{k}}^{2} - B_{\bf{k}}}} = - \omega_{4},
\end{eqnarray}
where
\begin{eqnarray}
&&A_{\bf{k}} = \frac{1}{2} \left( \epsilon_{\bf{k}}^{a 2} + \epsilon_{\bf{k}}^{b 2} + |\Delta_{a}|^{2} + |\Delta_{b}|^{2} + 2 V^{2} \right) \nonumber\\
  && B_{\bf{k}} \!=\! (\epsilon_{\bf{k}}^{a}\epsilon_{\bf{k}}^{b} \!-\! V^{2})^{2} \!+\! \epsilon_{\bf{k}}^{a 2} |\Delta_{a}|^{2} \!+\! \epsilon_{\bf{k}}^{b 2} |\Delta_{b}|^{2}
\nonumber \\ && \qquad
  \!+\! |\Delta_{a}||\Delta_{b}| \left[ |\Delta_{a}||\Delta_{b}| \!+\! 2 V^{2} \cos (\phi_{a} \!-\! \phi_{b}) \right]. \nonumber\\
\end{eqnarray}
where $\epsilon_{\bf{k}}^a=\epsilon_{0\bf{k}}^a-\mu$ and $\epsilon_{\bf{k}}^b=\epsilon_{0}+\alpha\epsilon_{0 \bf{k}}^a-\mu$.
The superconducting order parameters of the two-bands can be obtained self-consistently using the Fluctuation-Dissipation theorem \cite{tyablikov} or  by minimization of the ground state energy of the system.  These procedures also yield the equations for the number of particles. We obtain at zero temperature a set of coupled self-consistent equations given by,
\begin{eqnarray}
  && |\Delta_{a}| = J \sum_{\bf{k}} \left\{ \frac{|\Delta_{b}| \exp[- i (\phi_{a} \!-\! \phi_{b})]}{2 (\omega_{1}\! +\! \omega_{2})} \right.
  \nonumber \\
&& + \left.
\frac{|\Delta_{b}| \exp[-i (\phi_{a} \!-\! \phi_{b})] \left[\epsilon_{\bf{k}}^{b 2} \!+\! |\Delta_{a}|^{2} \right] \!+\! V^{2} |\Delta_{a}|}{2 \omega_{1} \omega_{2}(\omega_{1} + \omega_{2})}  \right\}, \nonumber \\
\label{Daa}
\end{eqnarray}

\begin{eqnarray}
 && |\Delta_{b}| = J \sum_{\bf{k}} \left\{ \frac{|\Delta_{a}| \exp[i (\phi_{a} \!-\! \phi_{b})]}{2 (\omega_{1}\! +\! \omega_{2})}\right. \nonumber \\
  && + \left. \frac{|\Delta_{a}| \exp[i (\phi_{a} \!-\! \phi_{b})] \left[\epsilon_{\bf{k}}^{a 2} \!+\! |\Delta_{b}|^{2} \right] \!+\! V^{2} |\Delta_{b}|}{2 \omega_{1} \omega_{2}(\omega_{1} + \omega_{2})}  \right\},\nonumber\\
\label{Dbb}
\end{eqnarray}

\begin{eqnarray}
  n_{a} &=&  \frac{1}{2}\sum_{\bf{k}} \left\{1 - \frac{\epsilon_{\bf{k}}^{a}}{(\omega_{1} + \omega_{2})}  \right.\nonumber \\
  && \quad + \left. \frac{\epsilon_{\bf{k}}^{b} V^{2} - \epsilon_{\bf{k}}^{a} (\epsilon_{\bf{k}}^{b 2} + |\Delta_{a}|^{2}) }{ \omega_{1} \omega_{2}(\omega_{1} + \omega_{2})}  \right\},
\label{na}
\end{eqnarray}

\begin{eqnarray}
  n_{b} &=&  \frac{1}{2} \sum_{\bf{k}} \left\{ 1 - \frac{\epsilon_{\bf{k}}^{b}}{(\omega_{1} + \omega_{2})} \right. \nonumber \\
  && \quad + \left. \frac{\epsilon_{\bf{k}}^{a} V^{2} - \epsilon_{\bf{k}}^{b} (\epsilon_{\bf{k}}^{a 2} + |\Delta_{b}|^{2}) }{ \omega_{1} \omega_{2}(\omega_{1} + \omega_{2})}  \right\}.
\label{nb}
\end{eqnarray}
Notice that due to hybridization, the conserved quantity in this case is the total number of particles, $n_{tot}=2(n_a+n_b)$, which is going to fix the chemical potential. The phases $\phi_a$ and $\phi_b$, so far remain arbitrary. Later on we will fix the difference between these phases.

In this paper we obtain solutions for these self-consistent equations in a two dimensional (2D) square lattice for several cases of physical interest. Since the chemical potential is also calculated self-consistently, we are not restricted to weak coupling and our results are valid for arbitrarily large coupling $J$.
We also present results for the density of states (DOS) for quasi-particles in bands $a$ and $b$ for $T=0 K$. These are obtained from the imaginary part of the Green's function for each band using,
\begin{eqnarray}
  \rho_{ab}(\omega)=\rho_a(\omega)+\rho_b(\omega)\,
\end{eqnarray}
where
\begin{eqnarray}
  \rho_a(\omega)=\frac{1}{\pi}\sum_{\mathbf{k}} \Im m \;G_{\bf{k}}^a(\omega)
\end{eqnarray}
\begin{eqnarray}
  \rho_b(\omega)=\frac{1}{\pi}\sum_{\mathbf{k}} \Im m \;G_{\bf{k}}^b(\omega)
\end{eqnarray}
where $G_{\bf{k}}^{i}(\omega)$ is the single particle Green's function of the $i^{th}$-band.

\section{Numerical Results}

In this Section, we obtain the zero temperature phase diagram of the two-band model, for different values of the interaction $J$, as a function of the occupation of the bands and the strength of the hybridization. Our aim when varying the latter is to simulate the effect of pressure in the system. The occupation naturally is changed by doping the material. Our results are obtained from a self-consistent solution of the coupled equations above, including the number equation. These solutions yield the order parameters $|\Delta_{a}|$, $|\Delta_{b}|$ and the chemical potential $\mu$ as a function of the parameters of the model, in particular the hybridization and the total number of electrons in the system. We consider mainly the case  the difference of phases, $\phi_a-\phi_b=0$ and the interaction $J$ is attractive. Superconducting solutions can also be obtained  taking $\phi_a-\phi_b=\pi$ and changing the sign of the interaction $J$ in  the gap equations \cite{iskin}. In solving the self-consistent equations, the sum over $k$ vectors in the square lattice is performed in the Brillouin zone using the method of special points from Chadi-Cohen~\cite{chadi}. Our numerical results shown below are obtained for inverted bands with a ratio of effective masses $\alpha=-0.7$. We renormalize all the physical parameters by the hopping term  $t=1$ of the $a$-band and  assume a bare band shift $\epsilon_0/t=2.0$.
\begin{figure*}[ht]
  \centering
  \includegraphics[width=0.9\textwidth]{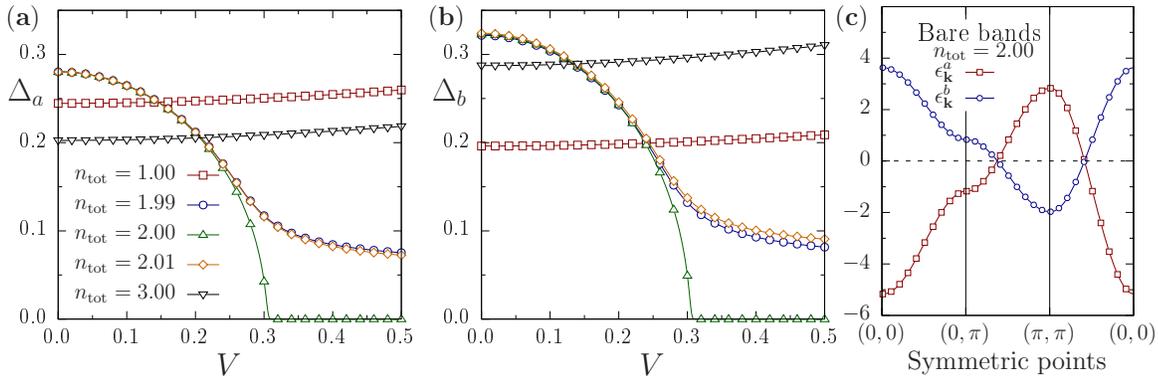}
  \caption{\label{figure01} Superconducting order parameters as a function of hybridization $V$ for $a$-band, {\bf (a)} $\Delta_a$, and $b$-band, {\bf (b)} $\Delta_b$, for different occupation numbers and $J=2.0$. {\bf (c)} Bare bands for $n_{\mathrm{tot}}=2.00$. All quantities are renormalized by the hopping $t$.}
\end{figure*}

Figures \ref{figure01}{\bf (a)} and \ref{figure01}{\bf (b)} show the superconducting order parameters  $|\Delta_{a}|$ and $|\Delta_{b}|$, respectively, as a function of hybridization $V$ for a coupling $J=2.0$.  For identical values of the occupation of the bands, $n_{\mathrm{tot}}$, both order parameters present similar variations with $V$.  This close relation between these quantities arises since the bands are homotectic, i.e., differ only by their effective masses and an energy shift. Besides, in the superconducting phase the order parameters are coupled by the Josephson term.  The bare bands, $\epsilon_{\mathbf{k}}^{a}$ and $\epsilon_{\mathbf{k}}^{b}$ are depicted in figure \ref{figure01}{\bf (c)}. In the absence of superconductivity and hybridization, these bands cross close to $(\pi,\pi)$. Any value of hybridization opens a gap, splitting them in two new bands.
We notice from the figures that in spite of their similar behavior, the  order parameters are not identical. $|\Delta_{a}|$ is favored for $n_{\mathrm{tot}} < 2.00$ and $|\Delta_{b}|$ is favored for $n_{tot} > 2.00$. From  now on without loss of generality, we will present  results only for $|\Delta_{a}|$.

As shown in figures~\ref{figure01}{\bf (a)} and \ref{figure01}{\bf (b)}  for $n_{\mathrm{tot}}=2.00$,  the order parameters vanish continuously at a zero temperature second order transition  for a critical value of the hybridization. For small deviations of this commensurate value of the total occupation, the order parameters instead of dropping to zero present a slow variation with no clear evidence for a quantum phase transition, as can be seen in the figures for  $n_{\mathrm{tot}}=1.99$ and $n_{\mathrm{tot}}=2.01$.
For $n_{\mathrm{tot}}=1.00$ and $n_{\mathrm{tot}}=3.00$, a radically different behavior is observed. In these cases, superconductivity is strengthened by increasing  hybridization. 
In order to understand this behavior we calculated several quantities for the two-band problem.

\begin{figure}[ht]
  \centering
  \includegraphics[width=0.9\columnwidth]{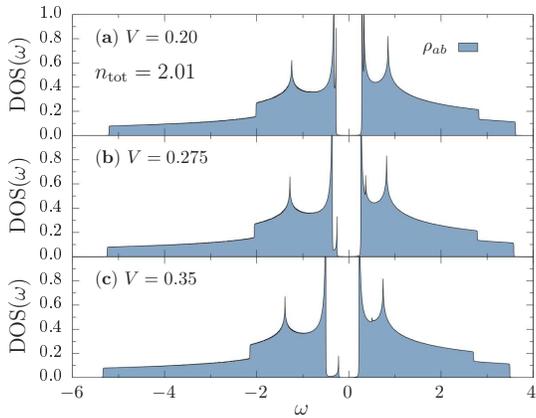}
  \caption{\label{figure02}    Density  of states for $n_{\mathrm{tot}}=2.01$ for different values of the hybridization $V$ and  $J=2.0$. As hybridization increases a sharp peak emerges from the lower part of the density of states and moves into the superconducting gap. This peak is associated with the regime of {\it tail-superconductivity} discussed in the text. For $n_{\mathrm{tot}}=1.99$ a similar effect is observed, with the peak emerging from the upper part of the density of states. All quantities are renormalized by the hopping $t$.}
\end{figure}

\begin{figure}[ht]
  \centering
  \includegraphics[width=0.9\columnwidth]{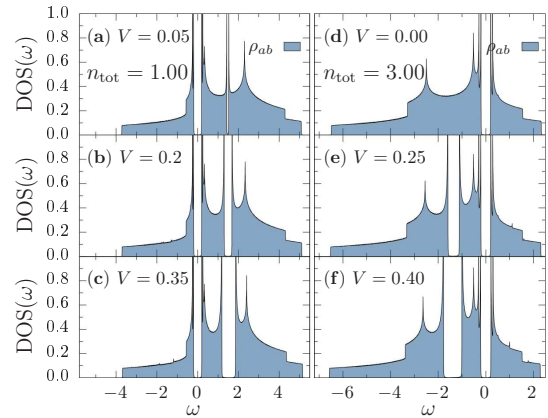}
  \caption{\label{figure03}  Density  of states for $n_{\mathrm{tot}}=1.00$ and $n_{\mathrm{tot}}=3.00$ and different values of hybridization $V$ for $J=2.0$. Notice the appearance of a hybridization gap, independent of the superconducting one, with increasing hybridization.  All quantities are renormalized by the hopping $t$.}
\end{figure}

The DOS for $n_{\mathrm{tot}}=2.01, 1.00, 3.00$  are shown in figures \ref{figure02} and \ref{figure03}, respectively.  Let us concentrate first on the occupations close to $n_{\mathrm{tot}}=2.00$. For $V \gtrsim 0.2$, the variations of $|\Delta_{a}|$ and $|\Delta_{b}|$ deviate significantly  from that for  $n_{\mathrm{tot}}=2.00$ as can be seen in figure~\ref{figure01}. This is accompanied by an interesting behavior in the DOS. A small part of the DOS splits from the main contribution and moves  inside the gap. For $n_{\mathrm{tot}}>2.00$ ($n_{\mathrm{tot}}<2.00$) this small peak splits from the upper (lower) band. As hybridization increases, the  gap continues to increase as  for $n_{\mathrm{tot}}=2.00$ (not shown), but for these occupations the small peak moves inside the gap towards the Fermi energy at $\omega=0$.  This contribution to the DOS is clearly associated with the persistence of superconductivity and the {\it tails} of the gap functions with increasing hybridization.  


Finally, notice that for $n_{\mathrm{tot}}=1.00$ and $n_{\mathrm{tot}}=3.00$ (figure \ref{figure03}), the hybridization and superconducting gaps are clearly distinct and superconductivity is not destroyed by hybridization.

\begin{figure}
  \centering
  \includegraphics[width=0.9\columnwidth]{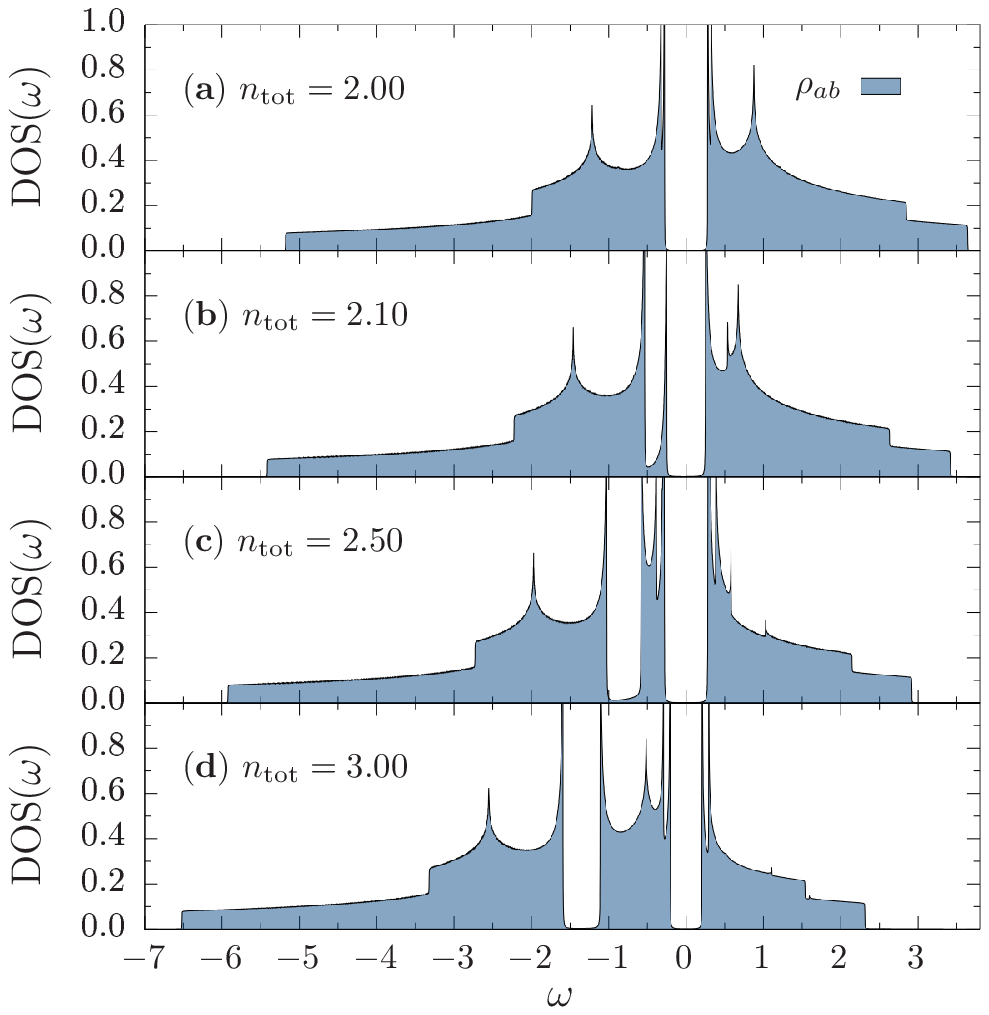}
  \caption{\label{figure04} The variation of the density of states for different values of $n_{\mathrm{tot}}$ and a fixed value of hybridization, $V=0.25$ for $J=2.0$. All quantities are renormalized by the hopping $t$.}
\end{figure}

\begin{figure}[ht]
  \centering
  \includegraphics[width=0.9\columnwidth]{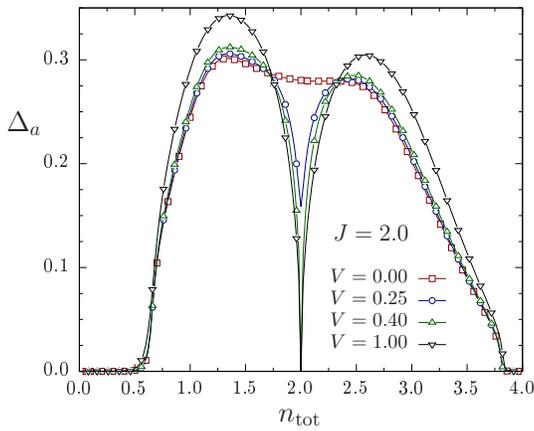}
  \caption{\label{figure05} Superconducting order parameter $\Delta_a$ as a function of the total number of particles $n_{\mathrm{tot}}$, and different values of hybridization $V$ for $J=2.0$. All quantities are renormalized by the hopping $t$.}
\end{figure}

\begin{figure}[ht]
  \centering
  \includegraphics[width=0.9\columnwidth]{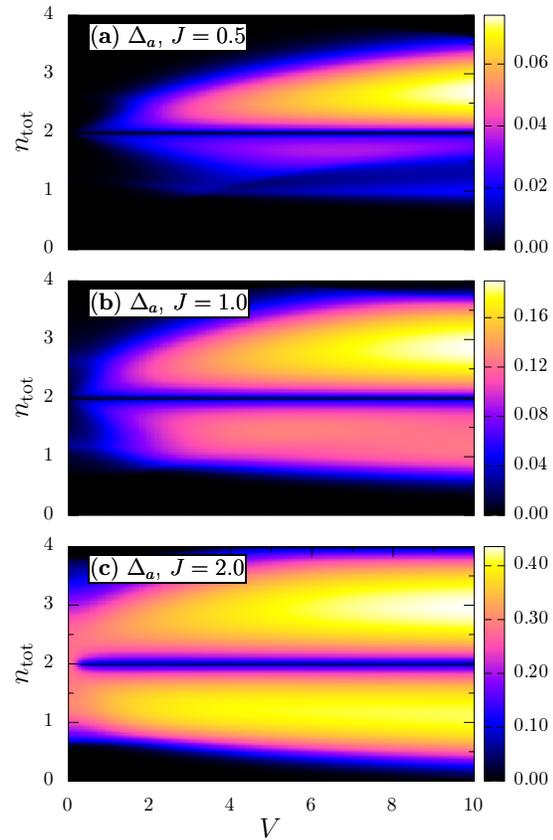}
  \caption{\label{figure06} Density plot showing the superconducting order parameter $\Delta_a$ (color) as a function of total number of particles $n_{\mathrm{tot}}$, and the hybridization $V$ for three values of $J$. All quantities are renormalized by the hopping $t$.}
\end{figure}

\begin{figure}[ht]
  \centering
  \includegraphics[width=0.9\columnwidth]{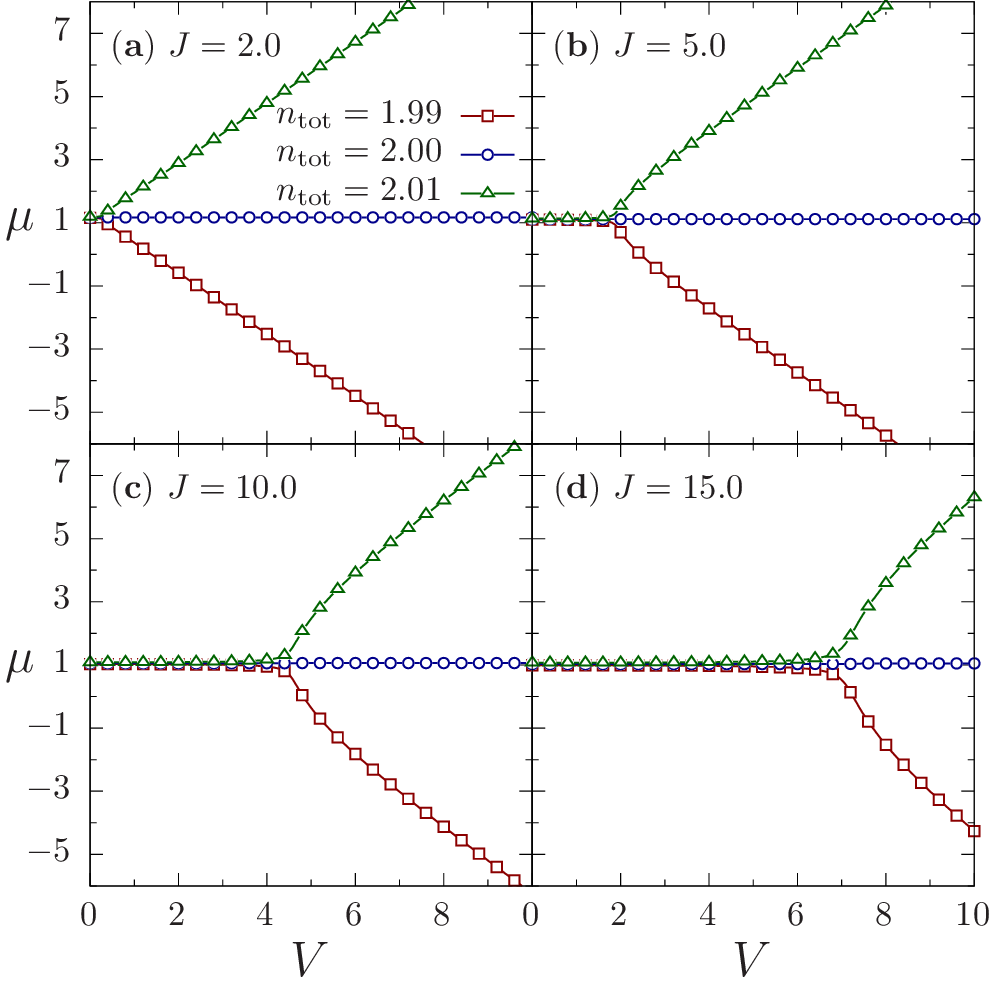}
  \caption{Chemical potential $\mu$ (Fermi level) as a function of hybridization $V$ for occupation number close to $n_{\mathrm{tot}}=2.00$ and different values of $J$. All quantities are renormalized by the hopping $t$.}
  \label{figure07}
\end{figure}

The variation of the density of states for different values of $n_{\mathrm{tot}}$ and a fixed value of hybridization,  $V=0.25$, is shown in figure \ref{figure04}. The superconducting  gap is always centered  at $\omega=0$ while the hybridization gap moves towards lower energies as $n_{\mathrm{tot}}$ increases. The maximum of the superconducting order parameter   $|\Delta_{a}|$ ($|\Delta_{b}|$) is observed around $n_{\mathrm{tot}}=1.50$ ($n_{\mathrm{tot}}=2.50$) as we can see in figure \ref{figure05}($\mathbf{c}$) and in figure \ref{figure06}. These figures show the clear difference between the hybridization and superconducting gaps for occupations $n_{\mathrm{tot}}\ne 2.00$.

The chemical potential as function of hybridization is shown in figure \ref{figure07},  for $n_{\mathrm{tot}}=1.99$, $2.00$ and $2.01$  and different values of the interaction $J$. The range of $V$ and $J$ is extended beyond physical values  in order to show clearly the strong deviation of $\mu$ in the regime of {\it tail-superconductivity}. For the commensurate occupation $n_{\mathrm{tot}}=2.00$ the chemical potential is independent of $V$ and remains fixed at the Fermi energy. However for small deviations of this value, as soon as the system enters in the {\it tail-regime} with increasing hybridization, the chemical potential deviates from the constant value for $n_{\mathrm{tot}}=2.00$. This deviation is associated with the appearance of  both, the tail regime, as observed in the superconducting order parameter as a function of $V$ and that of the small sharp peak inside the gap in the density of states. 

\subsection{The strong interaction limit}

\begin{figure}[ht]
  \centering
  \includegraphics[width=0.9\columnwidth]{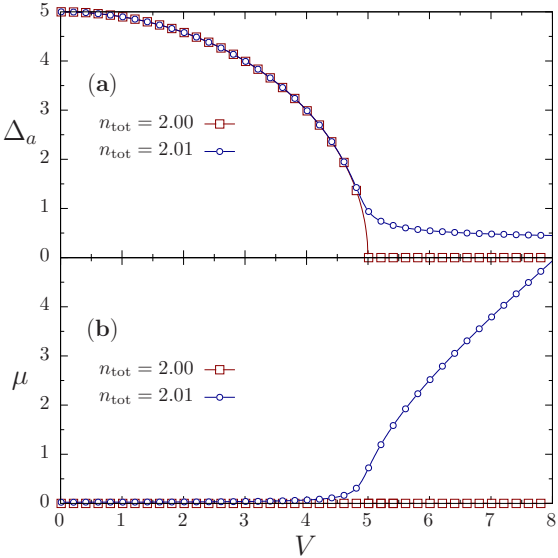}
  \caption{\label{figure08} {\bf (a)} Self-consistent solution \ref{s++} for the gap as a function of hybridization $V$ for $J/t=10.0$ and $n=2.00$, and $n=2.01$. {\bf (b)} Self-consistent solution of \ref{s++} for the chemical potential  as a function of hybridization $V$ for $J/t=10.0$ and $n=2.00$, and $n=2.01$. All quantities are renormalized by the hopping $t$.}
\end{figure}
In order to clarify the nature of the {\it tail-regime} of superconductivity, let us consider a strong-coupling expansion~\cite{Daniel15,Micnas1}  of the gap and number equations, (\ref{Daa}), (\ref{Dbb}), (\ref{na}) and (\ref{nb}) respectively. We consider both cases of $s^{++}$ and $s^{+-}$ symmetries that correspond to taking $\delta \phi = \phi_a - \phi_b$ equal  zero or $\pi$, respectively~\cite{fese,rafa1}. In the first case we get
\begin{eqnarray}
 && \frac{2}{J}=\frac{1}{\sqrt{(\mu+V)^2 + \Delta^2}+\sqrt{(\mu-V)^2+\Delta^2}}
  \nonumber \\
  &&\times\left[1+\frac{\mu^2+\Delta^2+V^2}{\sqrt{(\mu^2+\Delta^2-V^2)^2+4V^2 \Delta^2}}\right]\,, \nonumber \\
&&n-2=\frac{2 \mu}{\sqrt{(\mu+V)^2 + \Delta^2}+\sqrt{(\mu-V)^2+\Delta^2}}
  \nonumber \\
  &&\times \left[1+\frac{\mu^2+\Delta^2-V^2}{\sqrt{(\mu^2+\Delta^2-V^2)^2+4V^2 \Delta^2}}\right]  \nonumber \\
\label {s++}
\end{eqnarray}
for the gap and number equations. For simplicity we took $\epsilon_{0\bf{k}}^b=-\epsilon_{0\bf{k}}^a$ and $|\Delta_{a}| = |\Delta_{b}| = \Delta$. 
A solution of these self-consistent equations for $n$ equal or near $2$ is shown in figures~\ref{figure08}{\bf (a)} and \ref{figure08}{\bf (b)}. For $n=2.00$ the chemical potential remains fixed, at its weak coupling value as the hybridization increases. The superconducting order parameter however vanishes at a superconducting quantum critical point (SQCP) for a critical value of the hybridization. For small deviations of the commensurate value, i.e., for $n=2.01$, the chemical potential deviates from that of the commensurate value as $V$ increases. In this case there is no SQCP and a residual superconductivity persists even for large values of $V$. This type of behavior is reminiscent of the crossover BCS-BEC that occurs as the strength of the attractive interaction between the quasi-particles increases \cite{Chen}. Here, this crossover is induced by the hybridization.

For the case of $s^{+-}$ symmetry of the order parameter the gap and number  equations in the limit of  large interactions yield,
\begin{eqnarray}
\frac{1}{J}&=&\frac{1}{\sqrt{\mu^2 + \Delta^2} } \nonumber \\
n-2&=&\frac{ \mu}{\sqrt{\mu^2 + \Delta^2} }. \nonumber \\
\label {s+-}
\end{eqnarray}
These equations can be solved and we get,  $\Delta=J\sqrt{(n-1)(3-n)}$ and $\mu=J(n-2)$, independent of $V$.  In this case of repulsive interactions, superconductivity occurs only for $n>1$ and $n<3$ in a dome around $n=2$ at which occupation it is more robust, such that,  $\Delta$  and consequently $T_c$ are maximum. This is an important difference between the two cases of repulsive and attractive interactions. In the latter, superconductivity can arise for arbitrarily low or high band-fillings as can be checked by numerically solving equations~\ref{s++}.

\section{Discussion}

We have studied in this work how hybridization affects the superconducting properties of multi-band superconductors aiming to model the effect of pressure in these systems. We have obtained that this changes differently the superconducting behavior depending on the occupation of the bands. For occupations at the commensurate value of $n=2$, increasing hybridization $V(P)$ ($P$ is pressure) eventually destroys superconductivity at a SQCP for a critical value $V_c$. For fractional occupations of the bands hybridization has a distinct and multiple behavior. Very close to half-filling ($n=2$) it gives rise to a {\it tail regime} of superconductivity, where superconductivity is slowly suppressed but with no evidence of SQCP. This {\it tail-regime} superconductivity is associated with the appearance of a  sharp narrow peak in the density of states inside the superconducting gap, as shown in figure \ref{figure02}. The quasi-particles associated with this peak are the equivalent of strongly coupled pairs that appear in the BCS-BEC crossover, in this case induced by hybridizations, as we discuss below. The way hybridizations promotes this crossover is by directly influencing the chemical potential in the same way that increasing the strength of the interactions does, as is clearly shown in figure \ref{figure07}~\cite{nozieres,samelos}. Indeed  we see in these figures that as $V$ increases, the chemical potential starts to deviate from their weak coupling values. This occurs for the same values of $V$ for which the systems enters the {\it tail-regime} as observed in the behavior of the superconducting order parameter.

For arbitrary occupations, hybridization actually enhances superconductivity. We argue here that the effect of increasing hybridization in these cases is very similar to increasing the strength of the interactions responsible for superconductivity and that give rise to a BCS-BEC crossover in strong coupled superconductors. The most clear and direct evidences of this effect are the increase of the superconducting order parameter and the deviation of the chemical potential from its weak coupling values as hybridization increases. This hybridization promoted BCS-BEC crossover opens interesting possibilities of studying this phenomenon in condensed matter since the strength of the interactions is hard to control in these systems.

Our study has been carried out in the case of inverted bands and mostly for the case of $s^{++}$ symmetry, where the interactions responsible for superconductivity are attractive. The strong coupling analysis has been developed for both types of symmetry. We pointed out an important difference between these two cases, namely the absence of superconductivity in the dilute limit of electrons and holes for  $s^{+-}$ symmetry, with no occupation restriction in the $s^{++}$ symmetric problem.

Although the present work does not  address any particular type of systems, the results are sufficiently general to be relevant for mostly multi-band strongly correlated superconductors. We have identified a criterion for the existence of a SQCP induced by hybridization and a regime of tail superconductivity in inverted band systems. We have shown that pressure, i.e., hybridization can induce a BCS-BEC crossover in multi-bands systems even in the presence of moderate interactions. 

\section{Acknowledgements} 

MAC would like to thank Heron Caldas for useful discussions and the Brazilian Agencies, CAPES, CNPq and FAPERJ for partial financial support.

\end{document}